\documentclass[prapplied,longbibliography,reprint,superscriptaddress,amsmath,amssymb,aps]{revtex4-2}
\usepackage[utf8]{inputenc}
\usepackage{amsthm,amsfonts,braket}
\usepackage{placeins}[verbose]
\usepackage{tabularx, longtable, multirow}
\usepackage{graphicx}
\usepackage{bm,dsfont}
\usepackage[english]{babel}
\usepackage{comment}
\usepackage[dvipsnames]{xcolor}
\usepackage{booktabs}
\usepackage{siunitx}
\usepackage{mathtools}
\usepackage{appendix}
\usepackage{hyperref}
\usepackage[normalem]{ulem}
\usepackage[T1]{fontenc}
\usepackage{comment}
\usepackage{url}   

\newcommand{\ch}{\color{black}}

\newcommand{\cH}{{\cal H}}
\newcommand{\up}{{\uparrow}}
\newcommand{\down}{{\downarrow}}

\newcommand{\s}{\sigma}

\newcommand{\beginsupplement}{
        \setcounter{table}{0}
        \renewcommand{\thetable}{S\arabic{table}}
        \setcounter{figure}{0}
        \renewcommand{\thefigure}{S\arabic{figure}}
        \setcounter{equation}{0}
        \renewcommand{\theequation}{S\arabic{equation}}}

\begin{document}


\title{Spin-Energy Entanglement of a Time-Focused Neutron}
\thanks{This manuscript has been authored by UT-Battelle, LLC under Contract No. DE-AC05-00OR22725 with the U.S. Department of Energy. The United States Government retains and the publisher, by accepting the article for publication, acknowledges that the United States Government retains a non-exclusive, paid-up, irrevocable, world-wide license to publish or reproduce the published form of this manuscript, or allow others to do so, for United States Government purposes. The Department of Energy will provide public access to these results of federally sponsored research in accordance with the DOE Public Access Plan (\href{http://energy.gov/downloads/doe-public-access-plan}{http://energy.gov/downloads/doe-public-access-plan}).}

\author{J. C. Leiner}
\thanks{These authors contributed equally to this work.}
\email{Corresponding author: leinerjc@ornl.gov}
\affiliation{Neutron Technologies Division, Oak Ridge National Laboratory, Oak Ridge, TN 37830 USA}
\affiliation{Physik-Department, Technische Universit\"{a}t M\"{u}nchen, D-85748 Garching, Germany}
\affiliation{Heinz Maier-Leibnitz Zentrum (MLZ), Technische Universit\"at M\"unchen, D-85748 Garching, Germany}

\author{S. J. Kuhn}
\thanks{These authors contributed equally to this work.}
\affiliation{Neutron Technologies Division, Oak Ridge National Laboratory, Oak Ridge, TN 37830 USA}

\author{S. McKay}
\thanks{These authors contributed equally to this work.}
\affiliation{Department of Physics, Indiana University, Bloomington, IN 47405 USA}
\affiliation{Center for Exploration of Energy and Matter, Indiana University, Bloomington, IN 47408 USA}

\author{J. K. Jochum}
\affiliation{Heinz Maier-Leibnitz Zentrum (MLZ), Technische Universit\"at M\"unchen, D-85748 Garching, Germany}

\author{F. Li}
\affiliation{Neutron Technologies Division, Oak Ridge National Laboratory, Oak Ridge, TN 37830 USA}

\author{A.A.M. Irfan}
\affiliation{Department of Physics, Indiana University, Bloomington, IN 47405 USA}
\affiliation{Institute for Quantum Computing, University of Waterloo, Waterloo, N2L 3G1, ON, Canada}

\author{F.~Funama}
\affiliation{Neutron Technologies Division, Oak Ridge National Laboratory, Oak Ridge, TN 37830 USA}

\author{D.~Mettus}
\affiliation{Physik-Department, Technische Universit\"{a}t M\"{u}nchen, D-85748 Garching, Germany}
\affiliation{Heinz Maier-Leibnitz Zentrum (MLZ), Technische Universit\"at M\"unchen, D-85748 Garching, Germany}

\author{L. Beddrich}
\affiliation{Heinz Maier-Leibnitz Zentrum (MLZ), Technische Universit\"at M\"unchen, D-85748 Garching, Germany}

\author{C.~Franz}
\affiliation{J\"ulich Centre for Neutron Science JCNS, Forschungszentrum J\"ulich GmbH Outstation at MLZ FRM\,-\,II, D-85747 Garching, Germany}

\author{J.~Shen}
\affiliation{Laboratory for Neutron Scattering and Imaging, Paul Scherrer Institut, 5232, Villigen, Switzerland}

\author{S.~R.~Parnell}
\affiliation{Faculty of Applied Sciences, Delft University of Technology, Mekelweg 15, 2629 JB Delft, The Netherlands}

\author{R.~M.~Dalgliesh} 
\affiliation{ISIS, Rutherford Appleton Laboratory, Chilton, Oxfordshire, OX11 0QX, UK}

\author{M.~Loyd}
\affiliation{Neutron Technologies Division, Oak Ridge National Laboratory, Oak Ridge, TN 37830 USA}

\author{N.~Geerits}
\affiliation{Atominstitut, TU Wien, Stadionallee 2, 1020 Vienna, Austria}

\author{G.~Ortiz}
\affiliation{Department of Physics, Indiana University, Bloomington, IN 47405 USA}
\affiliation{Quantum Science and Engineering Center, Indiana University, Bloomington, IN 47408, USA}
\affiliation{Institute for Advanced Study, Princeton, NJ 08540, USA}

\author{C.~Pfleiderer}
\affiliation{Physik-Department, Technische Universit\"{a}t M\"{u}nchen, D-85748 Garching, Germany}
\affiliation{Heinz Maier-Leibnitz Zentrum (MLZ), Technische Universit\"at M\"unchen, D-85748 Garching, Germany}
\affiliation{Center for Quantum Engineering (ZQE), Technische Universit\"at M\"unchen, D-85748 Garching, Germany}
\affiliation{Munich Center for Quantum Science and Technology (MCQST), Technische Universit\"at M\"unchen, D-85748 Garching, Germany}

\author{R.~Pynn}
\affiliation{Department of Physics, Indiana University, Bloomington, IN 47405 USA}
\affiliation{Center for Exploration of Energy and Matter, Indiana University, Bloomington, IN 47408 USA}
\affiliation{Quantum Science and Engineering Center, Indiana University, Bloomington, IN 47408, USA}
\affiliation{Neutron Sciences Directorate, Oak Ridge National Laboratory, Oak Ridge, TN 37830 USA}

\date{\today}
\begin{abstract}
Intra-particle entanglement of individual particles such as neutrons could enable another class of scattering probes that are sensitive to entanglement in quantum systems and materials.
In this work, we present experimental results demonstrating quantum contextuality as a result of entanglement between the spin and energy modes (i.e., degrees of freedom) of single neutrons in a beam using a pair of resonant radio-frequency neutron spin flippers in the MIEZE configuration (Modulated IntEnsity with Zero Effort). We verified the mode-entanglement by measuring a Clauser-Horne-Shimony-Holt (CHSH) contextuality witness $S$ defined in the spin and energy subsystems, observing a clear breach of the classical bound of $|S| \leq 2$, obtaining $S = 2.40 \pm 0.02$.
These entangled beams could enable alternative approaches for directly probing dynamics and entanglement in quantum materials whose low-energy excitation scales match those of the incident entangled neutron.
\end{abstract}

\maketitle

\emph{Introduction.}---
The multitude of demonstrations of the physical limits of determinacy inherent to quantum superposition and its extension, quantum entanglement~\cite{PhysRevLett.128.160402}, all point directly to their fundamental utility~\cite{PhysRevLett.96.110404,PhysRevLett.118.060401}.
Quantum contextuality implies that even for observables at the same location in space and time (i.e., within a single particle \cite{azzini_review_2020}) there are still no hidden variables that determine observations \cite{RMP_QIS}.
In other words, the specific values for the observables are ``decided'' at the moment of measurement. Bell's theorem clarifies the way to rigorously distinguish between the case where particles hold information about their internal states or if those states are actually determined only at the instant of wave-function collapse. With carefully constructed experiments and closed loopholes supporting the latter case, it is becoming clear that quantum entanglement is an indispensable common element underlying fundamental physics, and may even be involved in the threads of spacetime itself \cite{Raamsdonk2010, PhysRevB.101.195134, PhysRevLett.119.080503, PhysRevA.89.052122}.

In general, there is an intense motivation to develop methods that probe entanglement in a more direct manner in order to provide independent evidence and facilitate observation of entanglement in materials \cite{Hauke_2016,Laflorencie_2016}. 
However, determining whether a system is entangled typically requires careful linking of theory and experiment. The methods investigated in this work are designed to be \textit{model agnostic}, which may allow direct extraction of information about quantum entanglement obtained from neutron spectroscopy experiments without the need for extensive modeling and prior knowledge of the sample's dynamics.
Instead of analyzing only the neutron spectrum scattered from a sample with conventionally prepared neutron beams, proof-of-concept efforts have shown how one can utilize neutron beams that are themselves prepared in an entangled state in order to probe entangled samples \cite{Irfan_2021}.

By performing calibrated polarization measurements using a neutron resonant spin-echo (NRSE) instrument \cite{Golub_1987}, we can obtain sufficient information on the correlations between the spin and energy observables of a single neutron to demonstrate the violation of a contextual Bell-like inequality \cite{shen2019,lu2019operator,kuhn2021}. Such a distinguishing demonstration is referred to as the construction and measurement of an \textit{entanglement witness}. The theory used to extract information about the entanglement of a system from witness measurements is already well established \cite{PhysRevB.103.224434, PhysRevLett.127.037201}. 
This type of measurement essentially amounts to entangling the different degrees of freedom of individual neutrons in a way that can be verified with established neutron scattering instrument techniques. For instance: the energy, spin, and path subsystems of an individual neutron can be entangled with each other using an NRSE  instrument \cite{shen2019}. Quantum contextuality is demonstrated by preparing and then experimentally measuring a Clauser-Horne-Shimony-Holt (CHSH) entanglement witness with the spin and path subsystems of the neutron \cite{hasegawa2003} or by preparing a Mermin witness with tripartite entanglement of spin, path, and energy \cite{hasegawa2010}. 

In this work, we demonstrate the entanglement of the energy and spin-distinguishable subsystems within a neutron as it is manipulated by a variant of NRSE spectroscopy called MIEZE (Modulated IntEnsity with Zero Effort). This entanglement is demonstrated by measuring a CHSH witness value of $2.40~\pm~0.02$.
This result is above the classical limit of 2 and corresponds to the expected value of a maximally entangled result of $2 \sqrt{2} \times C$ where $C = 85\%$ is the
MIEZE contrast (see Methods).
Such a measurement substantiates the viability and practicality of preparing an entangled neutron beam with a time-focusing condition in contrast to conventional NRSE.
In other words, this work {\ch (in contrast to previous work)} demonstrates that an established spectroscopy technique, with permanently installed incarnations such as the RESEDA beamline at FRM-II \cite{Franz2019,Franz_2019}, is intrinsically configured to exploit neutron spin-energy entanglement. {\ch Since the MIEZE technique already requires spin-energy correlations to achieve neutron measurements of dynamical excitations with \textit{both} high intensity and high resolution, it opens another potentially more effective and complementary avenue for intraparticle entanglement to serve as a probe for quantum materials, particularly in magnetic systems displaying a large degree of entanglement such as quantum spin liquids \cite{Khatua_2023}.}

\emph{Methods.}---
We first construct a particular CHSH witness using the spin ($s$) and energy ($e$) degrees of freedom of the neutron which are treated as two distinguishable subsystems leading to the tensor-product Hilbert state space \mbox{$\mathcal{H}=\mathcal{H}_{s}\otimes\mathcal{H}_{e}$} \cite{lu2019operator}. Both $\cH_s$ and $\cH_e$ describe two-level (i.e., qubit) subsystems: $\cH_s$ is the usual subspace of a nonrelativistic two-component spin-$1/2$ spinor and $\cH_e$ is the subspace spanned by two energy states.
The energy subspace can be described as two dimensional because the neutron can only access two energy states at each specific instance in time.
We denote the relative phases of the spin and energy states by $\alpha$ and $\gamma$, respectively.
We now define two pairs of observables $\sigma^s(\alpha_i)$ and $\sigma^e(\gamma_i)$ with $i~\in~\{1,2\}$ which act on the corresponding subsystems; these operators are associated with azimuthal angle $\alpha_i$ ($\gamma_i$) in the $x$-$y$ plane of the corresponding Bloch spheres:
\begin{subequations}
\begin{align}
    \s^{s}(\alpha_i) =&\cos{\alpha_i} \, \s^{s}_x+\sin{\alpha_i} \, \s^s_y \label{Eq: Def_Pauli_spin} \\
    \s^e(\gamma_i) =&\cos{\gamma_i} \, \s^e_x+\sin{\gamma_i} \, \s^e_y. \label{Eq: Def_Pauli_energy}
\end{align}
\end{subequations}
The projectors $P^s(\alpha_i)$ and $P^e(\gamma_i)$ for the observables $\s^s(\alpha_i)$ and $\s^e(\gamma_i)$ are defined, respectively, as
\begin{subequations}
\begin{eqnarray}
    P^s(\alpha_i)&=&\ket{\alpha_i}\bra{\alpha_i}, \ \quad \ket{\alpha_i}=\frac{\ket{\up}+e^{i \alpha_i}\ket{\down}}{\sqrt 2} \label{eq:spin proj}    \\
    P^e(\gamma_i)&=&\ket{\gamma_i}\bra{\gamma_i}, \ \quad \ket{\gamma_i}=\frac{\ket{E_+}+e^{i \gamma_i}\ket{E_-}}{\sqrt 2}.	\label{eq:energy proj}   
\end{eqnarray}
\end{subequations}
Next, we define the CHSH witness $S$ as 
\begin{equation} \label{Eq: CHSH Witness}
    S = E(\alpha_1, \gamma_1)+E(\alpha_1, \gamma_2)+E(\alpha_2, \gamma_1)-E(\alpha_2, \gamma_2),
\end{equation}
where the joint expectation values $E(\alpha_i,\gamma_j)$ for $i,j~\in~\{1,2\}$ are defined as
\begin{equation}
    E(\alpha_i, \gamma_j) = \bra{\psi}\sigma^s(\alpha_i) \sigma^e(\gamma_j)\ket{\psi}
\end{equation}
for a state $\ket{\psi} \in {\cal H}$.
By decomposing each observable into two projectors (geometrically represented by the antipodal points on the equator of the Bloch sphere), we can write each expectation value as
\begin{equation} \label{Eag}
    E(\alpha_i,\gamma_j) = \frac{\sum_{k,l} (-1)^{k + l} \, N(\alpha_i + k \pi,\gamma_j + l \pi)}{\sum_{k,l} N(\alpha_i + k \pi,\gamma_j + l \pi)}
\end{equation}
with $k,l \in \{0,1\}$. It follows that to determine each expectation value, measurements of at least four different phase-shift settings are needed, namely $\{N(\alpha_i~+~k\pi,\gamma_j~+~l\pi)\}$ with $k,l~\in~\{0,1\}$. 

No classical assignments of eigenvalues of observables by a local hidden variable theory can violate the CHSH inequality $|S|\leqslant 2$, but quantum mechanical expectations can \cite{Gisin_1991}. The maximum value for $S$ set by quantum mechanics is the Tsirelson bound of $2 \sqrt{2}$ \cite{cirel1980quantum}. Therefore, we have a straightforward criteria to detect the presence of quantum correlations, namely 
\begin{equation}
\begin{aligned}
    |S|& \leqslant 2 \quad \quad   \text{(classical statistics)}, \\
    |S|& \leqslant2\sqrt2  \quad  \text{(quantum statistics)}. \nonumber
\end{aligned}
\end{equation}
Any state violating the CHSH inequality is necessarily an entangled state in the spin and energy degrees of freedom. For the witness given in Eq. \eqref{Eq: CHSH Witness}, the maximum violation of the CHSH inequality occurs when $\alpha_1 + \gamma_1 = -\pi / 4$ and $\alpha_2 - \alpha_1 = \gamma_2 - \gamma_1 =\pi / 2$ \cite{lu2019operator}.

\begin{figure*}[t]
\includegraphics[width=\linewidth]{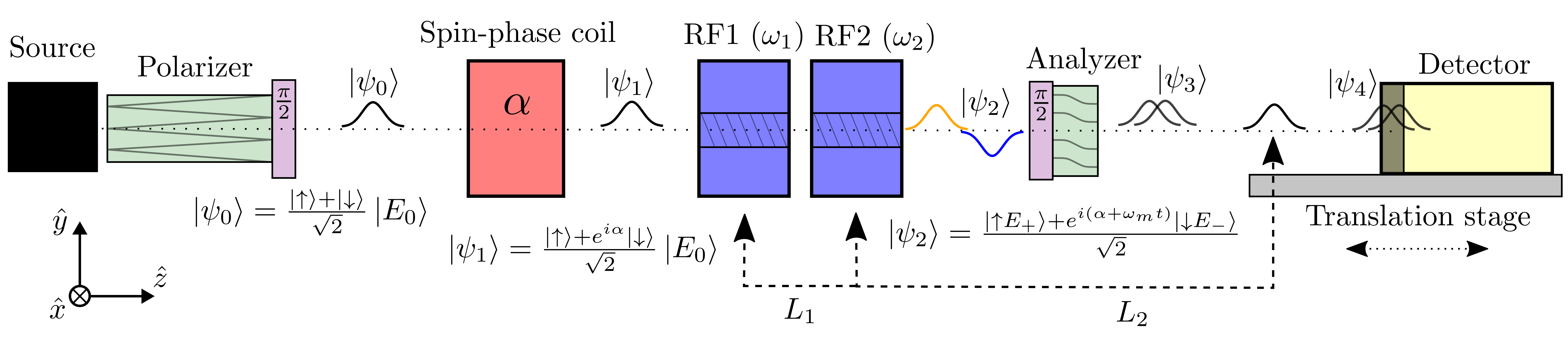}
\caption{\label{fig:2D experiment setup} 
Evolution of the total neutron wave function through the MIEZE instrument. After emerging from the polarizing V-cavity and first $\pi / 2$ flipper, the spin-phase coil tunes the spin phase of the neutron state $\ket{\psi_0}$ according to Eq. \eqref{eq:alpha}, resulting in the state $\ket{\psi_1}$. After the spin-phase coil, the neutron wave function is split both energetically and spatially by the first rf flipper (RF1), resulting in a maximally entangled Bell state \cite{Nielsen_2000}. After the second rf flipper (RF2) and second $\pi / 2$ flipper, the state is given by $\ket{\psi_2}$, which is still maximally entangled.
The analyzer acts like a projective measurement in the spin subsystem, producing an intensity oscillation in time, which is then measured by the detector. The detector is mounted on a longitudinally translating stage that adjusts the relative energy phase $\gamma$ given by Eq.~\eqref{gamma}. Additional weak magnetic guide fields {\ch in the $\hat{y}$ direction} are not shown.
The states $\ket{\psi_3}$ and $\ket{\psi_4}$ are both given by Eq. \eqref{eq_psi_3}.
{\ch The wavepacket sizes and spatial separations are greatly enhanced for clarity.}}
\end{figure*}

We now present a detailed description of the neutron state manipulations performed by a {\ch prototypical} MIEZE beamline as shown in Fig. \ref{fig:2D experiment setup}, thereby connecting the experimental measurement to the evaluation of the witness.
The standard MIEZE setup consists of two resonant rf neutron spin-flippers operated at different frequencies to create an intensity modulation of the neutron wave function after polarization analysis, resulting in a time-oscillating signal
\cite{1998Besenbock,Felber_1999}.
In the absence of additional magnetic fields, this beating frequency, commonly called the \textit{MIEZE frequency}, is given by
\begin{equation} \label{eq.genMIEZEfreq}
    \omega_m = 2 (\omega_2 - \omega_1),
\end{equation}
where $\omega_1$ ($\omega_2$) is the angular frequency of the first (second) rf flipper. 
Quantum mechanical treatments of the NRSE and MIEZE techniques have previously been described \cite{arend2011quantum,Keller2002}, and their relevant aspects are included later in this section. Additional details on the derivation of the entanglement witness and the necessary approximations are provided in the Supplemental Material \cite{SuppMat}. 

The experiment was performed on the CG-4B polarized test beamline at the High-Flux Isotope Reactor (HFIR) at Oak Ridge National Laboratory (ORNL), following initial measurements at FRM-II's RESEDA instrument~\cite{SuppMat}.
At CG-4B, the neutron beam is prepared by a silicon monochromator with wavelength 0.55~nm and bandwidth $\Delta\lambda/\lambda~\approx~0.2\%$ {\ch \footnote{Note that with a standard MIEZE beamline, the bandwidth is much higher, for example $\Delta\lambda/\lambda~\approx~12\%$ at the RESEDA beamline.}}.
The neutrons are then polarized using a V-cavity, ensuring that they have a well-defined initial spin state (96\% average polarization). They then pass through a $\pi/2$ flipper that initiates the neutron precession, creating a superposition of spin-up and spin-down states:
\begin{equation} \label{eq: E0 state}
   \ket{\psi_0} = \frac{\ket{\uparrow} + \ket{\downarrow}}{\sqrt{2}} \ket{E_0},
\end{equation}
where $E_0$ is the initial neutron energy {\ch and the quantization axis is chosen to be in the guide field direction (the $\hat{y}$ direction in Fig. \ref{fig:2D experiment setup}).}
After exiting the first $\pi /2$ flipper, the neutron travels through a static-field spin-phase coil, which produces a small magnetic field that is used to tune the neutron spin phase $\alpha$. The relative spin phase is given by the usual Larmor precession formula:
\begin{equation} \label{eq:alpha}
    \alpha = \frac{\gamma_n m \lambda}{h} \int d\ell \, \bm{B}(\ell),
\end{equation}
where $\gamma_n~>~0$ is the \emph{magnitude} of the neutron gyromagnetic ratio, $m$ the neutron mass, $\lambda$ the neutron wavelength, and $\int d\ell \, \bm{B}(\ell)$ the magnetic field integral experienced by the neutron along the path $\ell$ through the coil. The field integral contribution from the spin phase coil is well-approximated by $\int d\ell \, \bm{B}(\ell) \approx B L$ for a field region of length $L$ and uniform field strength $B$.
The input state to the first rf flipper up to a global phase factor is thus
\begin{equation} \label{eq: psi1 main text}
   \ket{\psi_1} = \frac{\ket{\uparrow} + e^{i \alpha}\ket{\downarrow}}{\sqrt{2}} \ket{E_0}.
\end{equation}

The neutron then proceeds through the two rf flippers, which in the MIEZE configuration are run at different frequencies. The first rf flipper (RF1) running at a frequency $\omega_1$ coherently changes the energy of the spin-up and spin-down states of the neutron. The change in energy corresponds to a change in velocity, and thus the up and down spin states separate spatially along the beamline ($\hat{z}$ in Fig. \ref{fig:2D experiment setup}).
The resonant spin-flip in the first rf flipper entangles the incident polarized neutron in spin and energy \cite{lu2019operator,SPONAR2010431}:
\begin{eqnarray}
    \ket{\up  E_0} \mapsto \ket{\down  E_-'},    \quad
    \ket{\down  E_0}\mapsto \ket{\up  E_+'}, \nonumber
\end{eqnarray}
where $E_{\pm}' = E_0 \pm \hbar \omega_1$.
Immediately after exiting the first rf flipper, the wave function of the neutron is then given up to a global phase factor by the maximally entangled Bell state \cite{Nielsen_2000}
\begin{equation} \label{Eq: Bell state}
    \ket{\psi_{\sf Bell}} = \frac{\ket{\up   E_+'}  +  e^{i (2 \omega_1 t - \alpha)} \ket{\down E_-'}}{\sqrt 2}.
\end{equation}

\begin{figure}[t]
\includegraphics[width=0.45\textwidth]{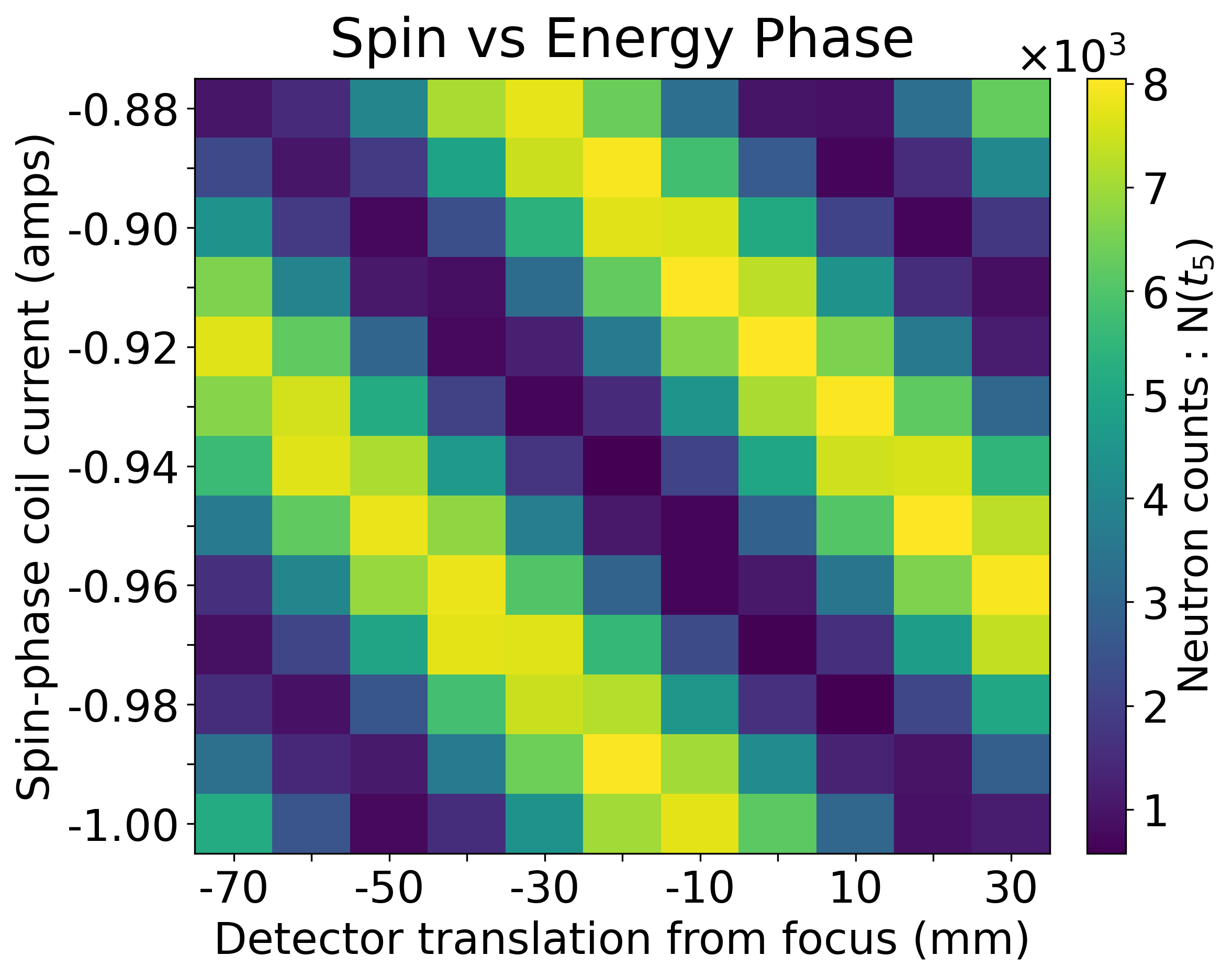}
\caption{\label{fig:2dscan} Relative intensities {\ch (see Ref. \cite{SuppMat} Sec. B for more details)} collected at specific detector translation positions and scanned through the spin-phase $\alpha$ at each detector position $\delta$ relative to the MIEZE focusing condition [see Eq.~\eqref{eq:FocusCondition}], which is proportional to the energy phase $\gamma$ [see Eq. \eqref{gamma}]. The frequencies of RF1 and RF2 were set at 45 kHz and 50 kHz, respectively. Negative values represent translation away from the analyzer. }
\end{figure}

\begin{figure}[t]
\includegraphics[width=0.49\textwidth]{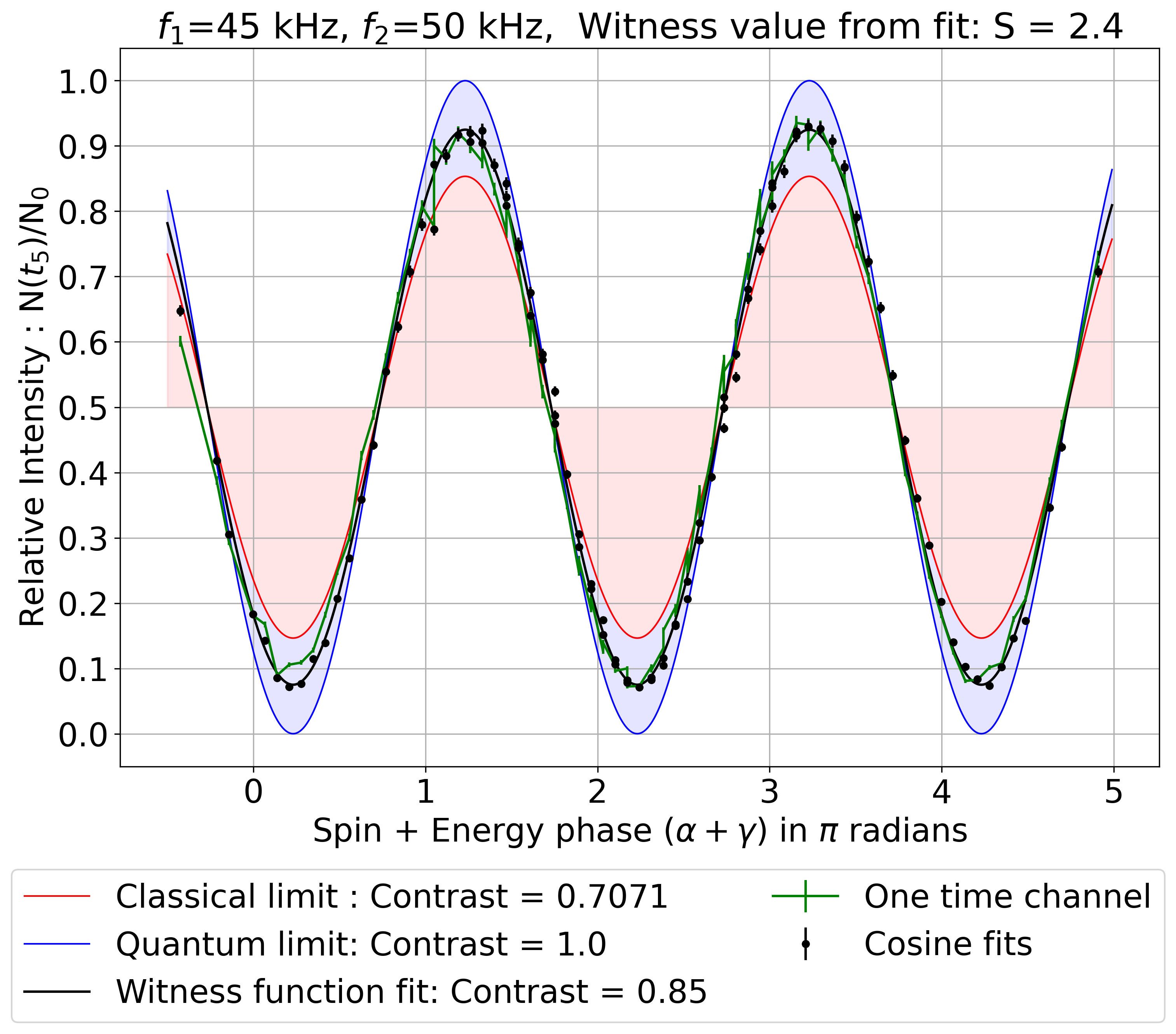}
\caption{\label{fig:cg4b witness} Data from Fig. \ref{fig:2dscan} fitted with $A + B \cos(\alpha + \gamma$) in accordance with Eq. \ref{cosinefit} and with the extracted parameters $A$ and $B$, the commensurate witness value is then calculated from Eqs. \eqref{Eag} and \eqref{Eq: CHSH Witness}. The blue colored area indicates the regime
where the witness value $S$ is strictly quantum, while the red area indicates values of $S$ that can arise from quantum or classical correlations. The error bars are shown for the measured intensities, which indicate the standard deviations resulting from counting statistics. }
\end{figure}

\noindent For more details on the origin of the relative phase in Eq.~\eqref{Eq: Bell state}, see Refs. \cite{Golub_1994,Ignatovich_2003}.
After exiting the second rf flipper (RF2) which again flips the neutron spin and changes the relative kinetic energy, the two states are longitudinally separated, with the lagging spin state having a greater energy; the neutron wave function  up to a global phase factor is given by
\begin{equation}
    \ket{\psi_2} = \frac{\ket{\uparrow E_+} + e^{i (\alpha + \omega_m t)} \ket{\downarrow E_-}}{\sqrt{2}},
\end{equation}
where $E_{\pm} = E_0 \pm \hbar (\omega_2 - \omega_1)$; this state is still a maximally entangled Bell state.
After the final $\pi/2$ flipper and the spin analyzer which {\ch transmits only the spin-up state}, the resulting neutron state becomes
\begin{equation} \label{eq_psi_3}
    \ket{\psi_3} = \frac{1}{2} \left[ \ket{E_+} + e^{i (\alpha + \omega_m t)} \ket{E_-} \right] \ket{\uparrow}.
\end{equation}
Note that the neutron state after the analyzer is no longer mode entangled between the spin and energy subsystems as the analyzer acts as the projection operator $P^s(\alpha=0)$ shown in Eq. \eqref{eq:spin proj}. Together, the spin-phase coil and analyzer act as a general projection operator $P^s(\alpha)$ for the spin subsystem.

In a typical MIEZE experiment, the detector is placed at the focusing point $z_f$ where the two wave packets recombine such that $|\braket{z_f | E_+}|^2~=~|\braket{z_f | E_-}|^2$. Complete recombination occurs only at a single point in space, but if the longitudinal beam coherence length $\beta_{\ell}~=~\lambda^2~/~\Delta \lambda$ is appreciably larger than the longitudinal spatial separation of the two spin states when the neutron is detected, then there is effectively a finite region of recombination \cite{Felber_1998,Arend_2004,PhysRevLett.50.563}. The region of recombination is determined by the MIEZE focusing condition given by 
\begin{equation} \label{eq:FocusCondition}
    \frac{L_1}{L_{2}} = \frac{\omega_2 - \omega_1}{\omega_1} + \frac{\gamma_n B L}{2\omega_1 L_{2}},
\end{equation}
where $L_1$ is the distance between the two rf flippers, $L_2$ the distance between the second rf flipper and the detector, and $B L$ the field integral due to the spin-phase coil \cite{Jochum2020b}.
This equation is simplified to neglect the length of the rf flipper itself and the static fields between the rf flippers (see Ref. \cite{Geerits2019} for a more complete focusing equation). Notice that the focusing condition does not depend on the wavelength of the neutron, which is key to the effectiveness of the MIEZE technique {\ch since more neutrons from the source can be utilized}.
If allowed to further propagate, the wave packets begin to spatially separate once more, resulting in the defocused state $\ket{\psi_4}$ shown in Fig.~\ref{fig:2D experiment setup} which is mathematically equivalent to Eq.~\eqref{eq_psi_3}.
When the focusing condition is applied, the relative energy phase at the detector position takes the simple form
\begin{equation} \label{gamma}
    \gamma = - \frac{m \lambda \omega_m}{h} \delta,
\end{equation}  
where $\delta$ is the displacement of the detector from the focusing point.
Therefore, at some point $\delta$ away from $z_f$ for a particular detector time channel $t_i$, we measure the {\ch neutron intensity as}
\begin{equation} \label{cosinefit}
\begin{aligned} 
    |\psi_{\delta}(t_i)|^2 =& \frac{1 + \cos(\alpha + \gamma + \omega_m t_i)}{2},
\end{aligned}
\end{equation}
which is equivalent to an $N(\alpha,\gamma)$ term in Eq.~\eqref{Eag}{\ch, again up to a global phase factor and fit parameters.} {\ch Taking $\gamma = 0$ at the focusing point, we must scan the detector longitudinally to adjust the energy phase (see Fig. \ref{fig:2D experiment setup}). Therefore, the location of the detector and kinematic recombination of the two states act as a general projection operator $P^e(\gamma)$ for the energy subsystem.}
From this analysis, we have shown that the spin phase coil, rf flippers, second $\pi/2$ flipper, analyzer, and detector mathematically represent a joint projective measurement with projection operator $P^s(\alpha)P^e(\gamma)$, allowing us to measure the CHSH witness defined in Eq.~\eqref{Eq: CHSH Witness}.

\emph{Results}---
The beamline was prepared in the MIEZE configuration using transverse rf flippers with high-temperature superconducting (HTS) coils generating the static magnetic field with HTS films at the boundaries to ensure sharp field transitions and improve field homogeneity \cite{dadisman2020design,Li_Dadisman_Wasilko_2020,mckay2024high}. The frequency of the first rf flipper was set at $f_1=45$~kHz and the second at $f_2=50$~kHz, resulting in a MIEZE frequency $\omega_m/(2\pi)$ = 10 kHz, {\ch which can be pushed into the MHz regime in the future \cite{Jochum2020, SuppMat, leiner2022miasans}}. The distance between the rf flippers was $L_1=85$~mm, which with a balanced guide field integral along the beamline [$BL=0$ in Eq.~\eqref{eq:FocusCondition}] sets  $L_2=765$~mm. The measurements consisted of independently scanning the spin and energy phases.
The spin phase $\alpha$ is adjusted by tuning the current in the HTS spin phase coil, which consisted of two rectangular coils in Helmholtz configuration surrounded by HTS films. The field integral of the coil changes by $BL~\approx~\SI{250}{\milli \tesla \milli \meter \per \ampere}$ per ampere of applied current. A $2\pi$ phase shift of $\alpha$ required a field integral of about $\SI{25}{\milli \tesla \milli \meter}$, corresponding to approximately a $\SI{0.1}{\ampere}$ change in current (see~Fig.~\ref{fig:2dscan}). Therefore, the spin-phase current was scanned from -1.00 to -0.88 A in $\SI{0.01}{\ampere}$ steps. The other guide fields through which the neutron passes {\ch together} contribute a small constant value to the spin phase; {\ch this additional global phase was} neglected as the guide fields were not changed during the experiment. 
Neutrons were counted with an Anger camera \cite{LOYD2024168871} mounted on a translation stage. A detector translation range of 70 mm covered a $2\pi$ energy phase shift range as shown in Fig.~\ref{fig:2dscan}.

The normalized intensities \cite{SuppMat} at the combined $\alpha$ and $\gamma$ phases was fit with a global cosine function $A + B $cos($\alpha + \gamma$) as shown in Fig.\ref{fig:cg4b witness}. The fitted parameters $A$ and $B$  were used to determine the expectation values defined by Eq.\eqref{Eag}. These expectation values $E(\alpha, \gamma)$ were applied to the CHSH witness (Eq.~\ref{Eq: CHSH Witness}) which yielded a value of 2.40 $\pm $0.02, well above the classical limit of 2; the observed witness value is also the maximum possible value with our 85\% MIEZE contrast, determined by the same fit. \\
\\

\emph{Conclusion}---
Based on recent theoretical and experimental work regarding mode-entangled neutron beams \cite{shen2019,lu2019operator,Irfan_2021,kuhn2021}, we have applied the theoretical procedure to rigorously construct a spin-energy entanglement witness using established MIEZE instrument configurations. The effectiveness of high energy-resolution neutron spin-echo spectroscopy techniques such as MIEZE fundamentally comes from labeling neutrons of varying energies with commensurate Larmor spin precessions. The procedure for witnessing spin-energy entanglement demonstrated here points to the utilization of the quantum properties of Larmor-labeled neutrons in inelastic neutron spectroscopy.  
High contrast and phase stability are the key requirements that must be ensured. The next steps will be to measure how suitable samples change the CHSH quantum contextuality witness value. Of note, with the MIEZE setup it is clear that the spin measurement at the analyzer is the point where the beam loses its spin-energy entanglement, and therefore the difference in signal with a sample placed before and after the analyzer would clarify the effects of an entangled neutron beam. Thus, this work represents a significant iteration toward a more direct and accessible method for probing entanglement in quantum materials. 

\emph{Acknowledgements}---
The authors gratefully acknowledge Doug Kyle, Kaleb Burrage, Lowell Crow, Goran Nilsen, and Mark Davenport for their assistance with experiments. Financial support by the German Bundesministerium für Bildung und Forschung (BMBF, Federal Ministry for Education and Research) through the Project No.\ 05K19W05 (`Resonante Longitudinale \mbox{MIASANS} Spin-Echo Spektroskopie an RESEDA') and Project No.\ 05K22WO2 (`Transportierbare Vorrichtung zur r\"aumlichen Intensit\"atsmodulation (SIM-Mode) von polarisierten Neutronenstrahlen am FRM-II') is gratefully acknowledged.

We gratefully acknowledge financial support by the Deutsche Forschungsgemeinschaft (DFG, German Research Foundation) under TRR 360 (project no. 492547816), the excellence cluster MCQST under Germany’s Excellence Strategy EXC-2111 (project no. 390814868), and the European Research Council (ERC) through Advanced Grant 788031 (ExQuiSid). This experiment was also attempted on the Larmor instrument at the ISIS Neutron and Muon source (UK) supported by a beamtime allocation RB2310511 from the Science and Technology Facilities Council \cite{UK_beamtime}.

This material is based upon work supported by the U.S. Department of Energy, Office of Science, Office of Workforce Development for Teachers and Scientists, Office of Science Graduate Student Research (SCGSR) program. The SCGSR program is administered by the Oak Ridge Institute for Science and Education for the DOE under contract number DE-SC0014664. This work was supported by the US Department of Energy (DOE), Office of Science, Office of Basic Energy Sciences, Early Career Research Program Award (KC0402010, ERKCSA4), under Contract No. DE-AC05- 00OR22725. This research used resources at the High Flux Isotope Reactor, a DOE Office of Science User Facility operated by the Oak Ridge National Laboratory. G.O. gratefully acknowledges support from the Institute for Advanced Study. S.J.K. was partially supported by and the rf flipper used in Oak Ridge was constructed and optimized by Department of Energy STTR grants DE-SC0021482, DE-SC0018453 and DE-SC0023624.\\

\bibliography{SEentanglementbib}
\clearpage
\newpage
\beginsupplement

\FloatBarrier
\section*{Supplemental Material}

\subsection{Evolution of the neutron wave packet}
In this section, we elaborate on the mathematical model used in the main text to describe the MIEZE technique. 
First, we elucidate the definitions of the energy states $\ket{E_0}$, $\ket{E_{\pm}'}$, and $\ket{E_{\pm}}$ used in the derivation of the contextuality witness, extending the results derived in \cite{Golub_1994,Ignatovich_2003} from plane waves to wave packets.
We also show how sufficient beam monochromaticity is the necessary requirement to treat the relative spin phase $\alpha$ and relative energy phase $\gamma$ as simple phases between the spinor components without needing to consider the particular structure and size of the neutron wave packet.
Finally, we show that the plane wave phase and spatially-separated wave packet treatments of the MIEZE technique can be derived from the wave packet model by using the integration technique of stationary phase where the value of the integral is dominated by the critical points of the rapidly-varying phase \cite{Dingle_1973,Tannor_2007}.

For simplicity, we only consider the longitudinal component of the neutron wave packet in position-space in our analysis. We also ignore the effects of the weak guide fields present throughout the instrument.
To start, we take as a fundamental assumption that the neutron beam consists of non-interacting, mutually incoherent wave packets. The wave function for a single incident neutron wave packet can then be represented by
\begin{equation}
    \braket{z | \psi_0} = \frac{1}{\sqrt{2} \mathcal{N}} \int_{\mathbb{R}} dk \, g(k - k_0) e^{i[kz - \omega(k)t]} \begin{pmatrix}
           1 \\
           1
    \end{pmatrix},
\end{equation}
where $g(k - k_0)$ is an arbitrary square-integrable function strongly-peaked at $k_0$ which represents the shape of the wave packet at time $t = 0$, $\omega(k) = \hbar k^2 / (2 m)$ the usual dispersion relation for a free particle, and $\mathcal{N}$ a potentially complex normalization which depends on the explicit shape of the wave packet. 
By inspection, the wave function of the state given in Eq. \eqref{eq: E0 state} in the main text is given by
\begin{equation}
    \braket{z | E_0} = \frac{1}{\mathcal{N}} \int_{\mathbb{R}} dk \, g(k - k_0) e^{i[kz - \omega(k)t]}.
\end{equation}

Next, assuming that $|\bm{\mu} \cdot \bm{B}| / E_0 \ll 1$ where $-\bm{\mu} \cdot \bm{B}$ is the magnetic component of the neutron Hamiltonian in the static field $\bm B$ and $E_0$ is the initial neutron energy, the wave function after the spin phase coil is well-approximated by
\begin{equation} \label{eq: wave packet psi1 gen}
    \braket{z | \psi_1} = \frac{1}{\sqrt{2} \mathcal{N}} \int_{\mathbb{R}} dk \, g(k - k_0) e^{i[kz - \omega(k)t]} 
     \begin{pmatrix}
           e^{-i \alpha(k) /2} \\
           e^{i \alpha(k) /2}
    \end{pmatrix},
\end{equation}
where we have explicitly kept the $k$ dependence in the spin phase, so
\begin{equation} \label{eq:gen alpha}
    \alpha(k) = \frac{2 \pi m \gamma B L}{h k},
\end{equation}
where $\alpha(k_0) = \alpha$ from Eq. \eqref{eq:alpha} in the main text.

Evaluating the integral of Eq. \eqref{eq: wave packet psi1 gen} via the method of stationary phase, we find that the neutron wave packet is now two-peaked:
\begin{equation}
    z_p = \frac{\hbar k_0}{m} t \mp \frac{\alpha}{2 k_0},
\end{equation}
with each peak $z_p$ corresponding to the spin-up ($-$ sign) or spin-down ($+$ sign) component of the neutron spinor. Note that the quantity $\hbar k_0 / m$ is the group velocity of the incident neutron wave packet.
This longitudinal splitting is an example of the single-particle Stern-Gerlach effect~\cite{Gerlach_Stern_1922,Sherwood_1954}.
For a sufficiently broad wave packet [i.e., a sufficiently sharp $g(k-k_0)$ distribution] with an intrinsic longitudinal coherence length $\Delta_{\ell}$, we can approximate this two-peaked wave function as a single-peaked wave function with a relative phase between the two spinor components.

Although the value of the intrinsic coherence length is unknown, the longitudinal beam coherence length $\beta_{\ell}$ is an upper bound of the intrinsic coherence length \cite{kuhn2021}. Therefore, writing $\Delta_{\ell} = \kappa \beta_{\ell}$ with $\kappa \geq 1$ some constant, a necessary condition to treat the split wave packet as a single-peaked wave packet with a relative phase between the two spinor components is
\begin{equation} \label{eq:alpha condition}
    \frac{|\alpha|}{k_0} \ll \Delta_{\ell} = \kappa \frac{\lambda^2}{\Delta \lambda} \Rightarrow
    \frac{|\alpha|}{2 \pi} \ll \kappa \frac{\lambda}{ \Delta \lambda}.
\end{equation}
Therefore, with the conservative estimate $\kappa = 1$, the condition reduces to the ``total number of precessions'' $N_{\alpha} = |\alpha| / (2 \pi)$ must be much less than the inverse of the wavelength bandwidth. For the monochromator at CG-4B, an order of magnitude difference corresponds to a limit of $N_{\alpha} \lesssim 50$.
This condition was satisfied for all measurements in this experiment, allowing us to write Eq. \eqref{eq: wave packet psi1 gen} modulo a global phase factor as
\begin{equation} \label{eq: wave packet psi1}
    \braket{z | \psi_1} = \frac{1}{\sqrt{2} \mathcal{N}} \int_{\mathbb{R}} dk \, g(k - k_0) e^{i[kz - \omega(k)t]} \begin{pmatrix}
           1 \\
           e^{i \alpha}
    \end{pmatrix},
\end{equation}
which agrees with Eq. \eqref{eq: psi1 main text} in the main text.

Now we consider the action of the first rf flipper on the neutron state. Extending the plane wave result derived in \cite{Golub_1994}, we find
\begin{equation}
\begin{aligned}
    \braket{z | \psi_{\mathrm{Bell}}} =& \frac{-i}{\sqrt{2} \mathcal{N}} \int_{\mathbb{R}} dk \, g(k - k_0) \times \\
    &\begin{pmatrix}
        e^{i \left[ \left(k + \frac{m \omega_1}{\hbar k} \right)z - (\omega(k) + \omega_1)t \right]} \\
        e^{-i \alpha} e^{i \left[ \left(k - \frac{m \omega_1}{\hbar k} \right)z - (\omega(k) - \omega_1)t \right]}
    \end{pmatrix} \\
    =& \frac{1}{\sqrt{2} \mathcal{N'}} \int_{\mathbb{R}} dk \, g(k - k_0) \times \\
    &\begin{pmatrix}
        e^{i \left[ \left(k + \frac{m \omega_1}{\hbar k} \right)z - \omega(k) t \right]} \\
        e^{-i \alpha} e^{i 2 \omega_1 t} e^{i \left[ \left(k - \frac{m \omega_1}{\hbar k} \right)z - \omega(k) t \right]}
    \end{pmatrix},
\end{aligned}
\end{equation}
where we have absorbed the additional phases irrelevant to our analysis into the new normalization $\mathcal{N'}$.
Therefore, we can define wave functions of the the energy state $\ket{E_{\pm}'}$ as
\begin{equation}
    \braket{z | E_{\pm}'} = \frac{1}{\mathcal{N'}} \int_{\mathbb{R}} dk \, g(k - k_0) e^{i \left[ \left(k \pm \frac{m \omega_1}{\hbar k} \right)z - \omega(k) t \right]},
\end{equation}
where $E_{\pm}' = E_0 \pm \hbar \omega_1$.
Using the method of stationary phase again and assuming $\hbar \omega_1 / E_0 \ll 1$, we see that peaks of the split neutron wave packet now follow the trajectories
\begin{equation}
    z_p \approx \frac{\hbar k_0}{m} \left( 1 \pm \frac{m \omega_1}{\hbar k_0^2} \right) t,
\end{equation}
where the spin-up (spin-down) component corresponds to the $+$ sign ($-$ sign). Notice that the difference in group velocities between the two peaks of the wave packet is given by $2 \omega_1 / k_0$, so the two wavepackets are longitudinally separated by the first rf flipper; the total spatial separation is determined by the flight path length $L_1$ between the the two flippers.

Finally, we consider the action of the second rf flipper which again flips the neutron spin and changes the relative energy, resulting in the neutron wave function
\begin{equation} \label{eq: psi2 gen}
\begin{aligned}
    \braket{z | \psi_2} =& \frac{-i}{\sqrt{2} \mathcal{N'}} \int_{\mathbb{R}} dk \, g(k - k_0) \times \\
    &\begin{pmatrix}
        e^{i \left[ \left(k + \frac{m (\omega_2 - \omega_1)}{\hbar k} \right)z - (\omega(k) + \omega_2)t - \frac{m \omega_1 L_1}{\hbar k} \right]} \\
        e^{i \alpha} e^{-i 2 \omega_1 t} e^{i \left[ \left(k - \frac{m (\omega_2 - \omega_1)}{\hbar k} \right)z - (\omega(k) - \omega_2)t + \frac{m \omega_1 L_1}{\hbar k} \right]}
    \end{pmatrix} \\
    =& \frac{1}{\sqrt{2} \mathcal{N''}} \int_{\mathbb{R}} dk \, g(k - k_0) e^{i[kz - \omega(k)t]} \times \\
    &\begin{pmatrix}
        e^{i \left[\frac{m (\omega_2 - \omega_1)}{\hbar k} z - \frac{m \omega_1 L_1}{\hbar k} \right]} \\
        e^{i \alpha} e^{i \omega_m t} e^{-i \left[\frac{m (\omega_2 - \omega_1)}{\hbar k} z - \frac{m \omega_1 L_1}{\hbar k} \right]}
    \end{pmatrix},
\end{aligned}
\end{equation}
where we have again absorbed the irrelevant factors into the new normalization $\mathcal{N''}$.
Therefore, the wave functions of the energy kets $\ket{E_{\pm}}$ are defined as
\begin{equation}
\begin{aligned}
    \braket{z | E_{\pm}} =& \frac{1}{\mathcal{N''}} \int_{\mathbb{R}} dk \, g(k - k_0) e^{i[kz - \omega(k)t]} \times \\
    & e^{\pm i \left[\frac{m (\omega_2 - \omega_1)}{\hbar k} z - \frac{m \omega_1 L_1}{\hbar k} \right]},
\end{aligned}
\end{equation}
where $E_{\pm} = E_0 \pm \hbar (\omega_2 - \omega_1)$.
Evaluating the integral in Eq. \eqref{eq: psi2 gen} via stationary phase assuming that $\hbar \omega_i/E_0 \ll 1$ for $i \in \{1,2\}$, we find that the two peaks of the neutron wave packet follow the trajectories
\begin{equation}
\begin{aligned}
    z_p &\approx \frac{\hbar k_0}{m} \left( 1 \pm \frac{m \omega_m}{2 \hbar k_0^2}\right) t \mp \frac{m \omega_1 L_1}{\hbar k_0^2}. 
\end{aligned}
\end{equation}
From the above equation, we see that the two peaks converge at a point $L_2 = 2 \omega_1 L_1 / \omega_m$, which is consistent with the focusing condition given in Eq. \eqref{eq:FocusCondition} in the main text when treating the contribution from the spin phase coil as a simple relative phase between the spinor components; to obtain Eq. \eqref{eq:FocusCondition} from the main text exactly, the form of spin phase $\alpha(k)$ given in Eq. \eqref{eq:gen alpha} must be used.

To define the energy phase $\gamma$, we consider the neutron state given in Eq. \eqref{eq: psi2 gen} at a point $z~=~L_2~+~\delta$. With the same assumptions used when deriving the conditions on the relative spin phase in Eq. \eqref{eq:alpha condition}, we find that the necessary condition to treat the energy phase as a simple relative phase between the spinor components is
\begin{equation} \label{eq: gamma bound}
    \frac{m}{h k_0} \omega_m \delta = \frac{| \gamma |}{2 \pi} \ll \kappa \frac{\lambda}{\Delta \lambda},
\end{equation}
which was also satisfied for all measurements in this experiment. Therefore, with these assumptions, the relative energy phase $\gamma$ at the point $\delta$ away from the focusing point can be expressed as 
\begin{equation}
    \gamma(k) = -\frac{2 \pi m}{h k} \omega_m \delta,
\end{equation}
where $\gamma(k_0) = \gamma$ as defined in Eq. \eqref{gamma} in the main text, resulting in Eq. \eqref{cosinefit} in the main text for the observed signal at the detector for a single neutron.

As an aside, we note that one could also tune the relative energy phase without moving the detector by changing the relative frequency between the two rf flippers. In that case, using a similar analysis to the one presented above, the relative energy phase at the focusing point would be given by $\gamma = -2 (\delta \omega) t$ where $\delta\omega$ is the deviation from the MIEZE frequency~$\omega_m$.

\clearpage
\newpage

\subsection{Data collection and fitting}
\begin{figure}[ht]
\includegraphics[width=0.49\textwidth]{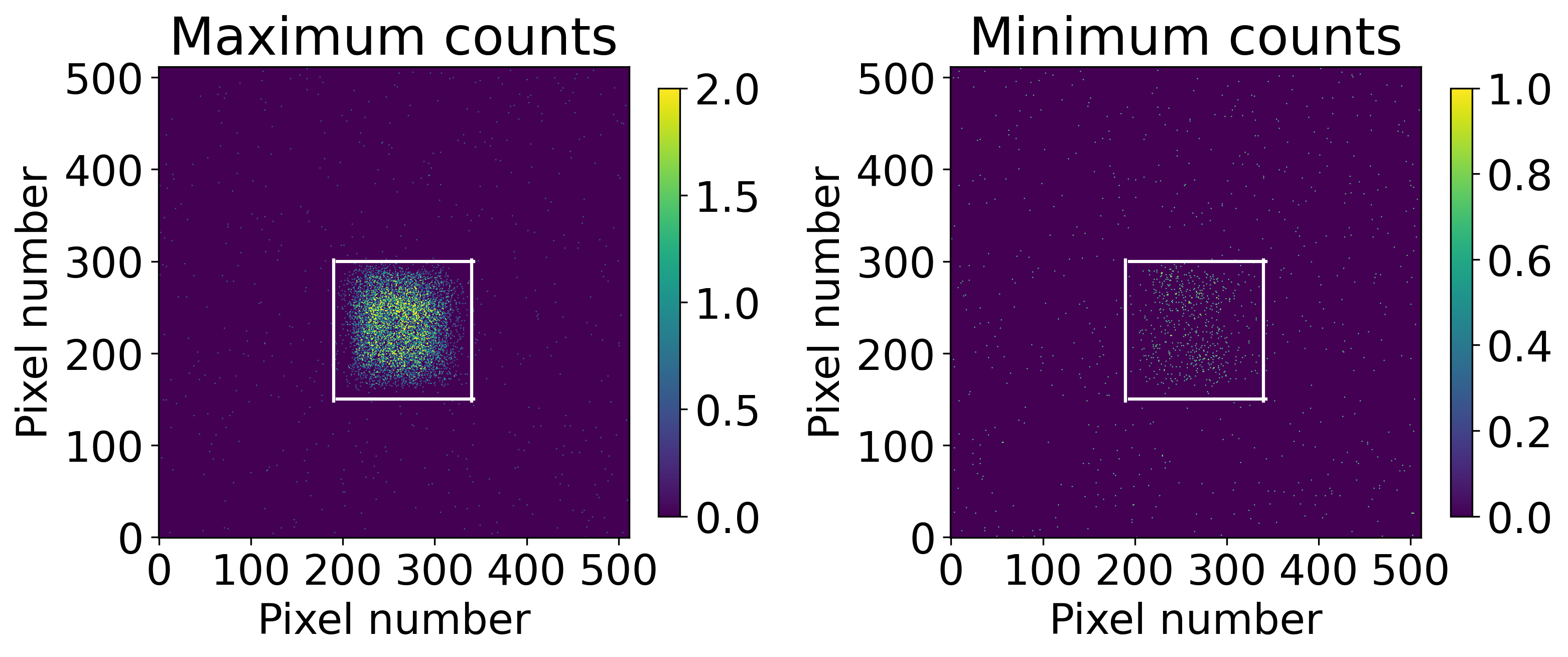}
\caption{\label{tof_images} The background-subtracted detector images that depict the highest and lowest intensity levels in a single period of oscillation on the Anger camera \cite{LOYD2024168871}. The neutrons were counted only in the direct beam region of interest (ROI) outlined by the white squares with edges 150~pixels in length ($\sim$3.5~cm$^2$). To convert to relative intensity used in Figs. \ref{fig:cg4b witness} and \ref{fig:cg4b witness supp}, the maximum and minimum counts in the ROI were added together: $N_0=8600$.}
\end{figure}

\FloatBarrier
Here we provide further details on the collection of the data points presented in Figures \ref{fig:2dscan} and \ref{fig:cg4b witness}. Fig. \ref{tof_images} shows the representative detector images at the maximum and minimum intensity of the MIEZE oscillation.

\begin{figure}[ht]
\includegraphics[width=0.49\textwidth]{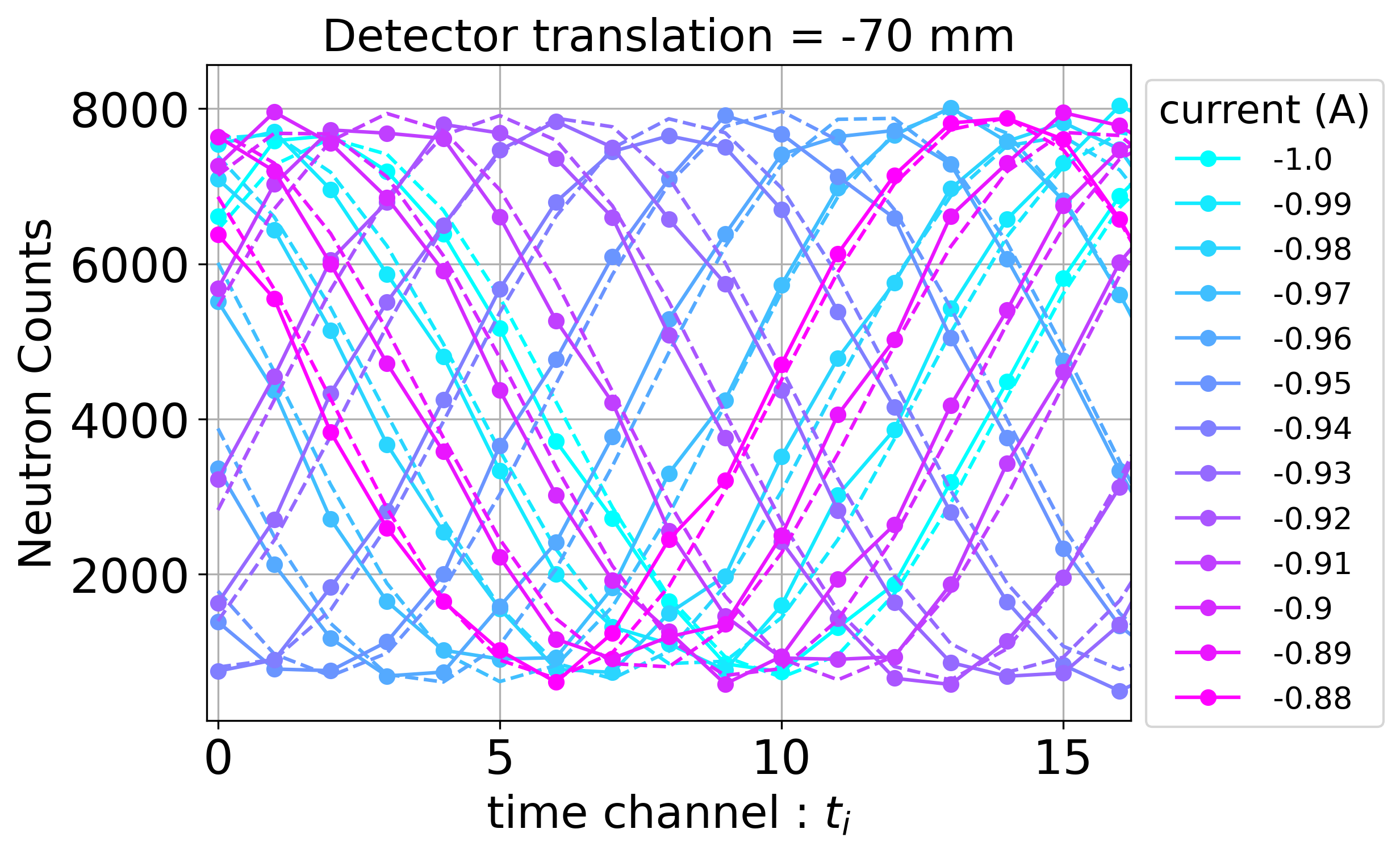}
\caption{\label{sample_data} An example of the data collected at a specific detector position, showing the neutron counts collected in each time channel at the specified currents mentioned in the Results section. The dashed lines show the respective cosine fits to the data at each spin-phase coil current.}
\end{figure}

Fig. \ref{sample_data} shows single periods of the MIEZE oscillation with the counts collected over the region of interest displayed in Fig. \ref{tof_images}. The phase shift is clearly well controlled, in this case using the spin-phase coil current. Data points in Fig.~\ref{fig:2dscan} and those labeled ``One time channel" in Fig. \ref{fig:cg4b witness} were taken from time channel number 5, $t_5$, an arbitrary choice. To make use of the data points in every time channel, each cosine curve was fitted, and the data points labeled ``Cosine fits" in Fig. \ref{fig:cg4b witness} were obtained from $t_5$ of these curve fits.

\subsection{Data from Timepix3 camera system}

\begin{figure}[hb]
\includegraphics[width=0.49\textwidth]{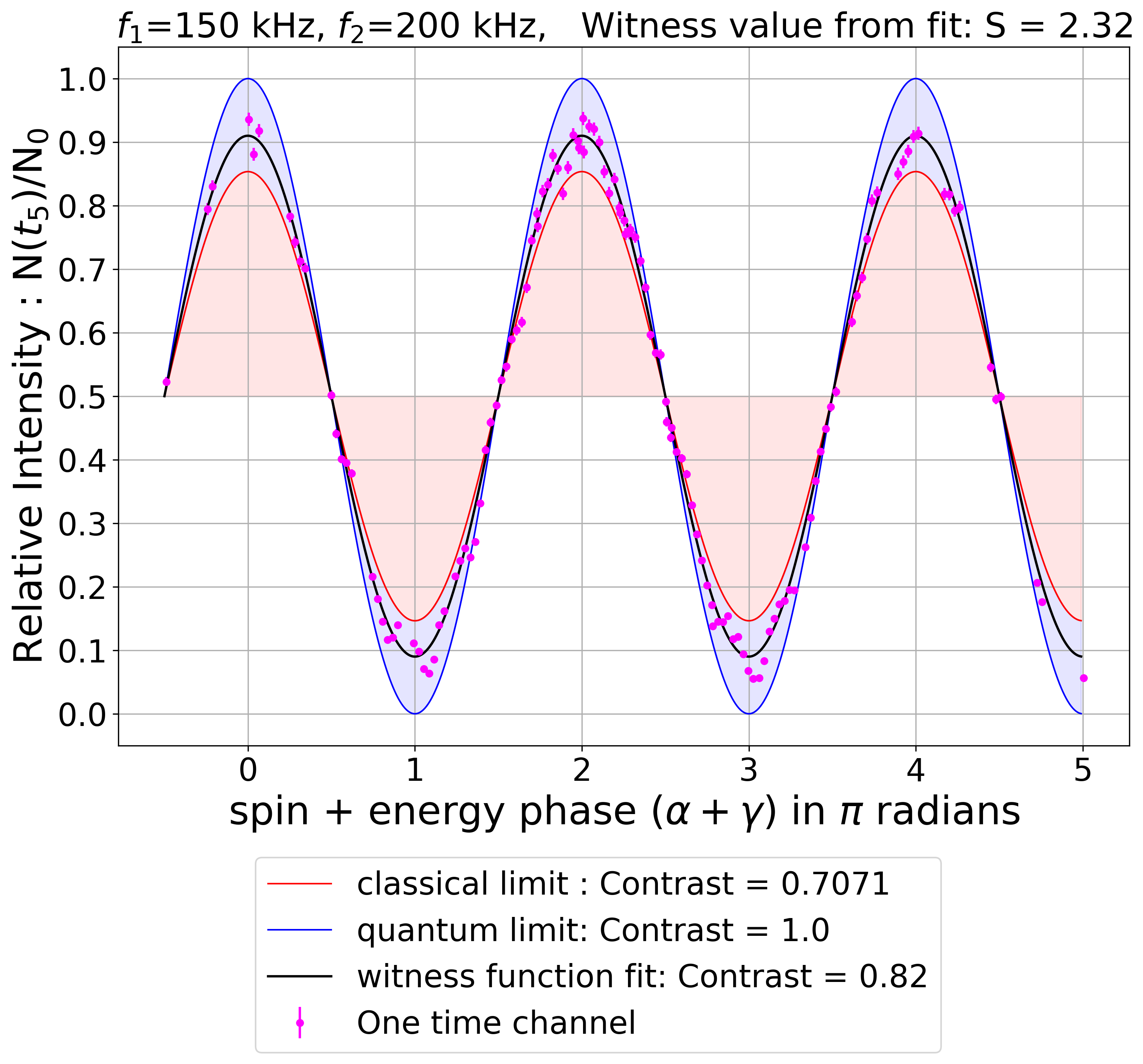}
\caption{\label{fig:cg4b witness supp} Relative intensities collected at specific detector translation positions and scanned through the spin phase $\alpha$ at each of these settings using the Timepix3 camera system. Data are fitted with the function $A + B\cos(\alpha + \gamma$) according to Eq. \eqref{cosinefit} in the main text. With the extracted parameters $A$ and $B$, the commensurate witness value is then calculated from Eqs. \eqref{Eag} and \eqref{Eq: CHSH Witness} in the main text.
Error bars are shown for measured intensities, which indicate standard deviations resulting from counting statistics. }
\end{figure}

The witness measurement was repeated using a higher frequency in the rf flippers. The frequencies were set at 150 kHz and 200 kHz, leading to $\omega_m/(2\pi)$~=~100 kHz. The spin phase coil was set at $\SI{1.38}{\ampere}$  (where the contrast was found to be highest at 82\%) and scanned over a range of $\SI{0.12}{\ampere}$; the detector position was scanned by 10 mm to modify the energy phase. Here, the detector scan range is ten times smaller than in the previous case, since the difference in frequency is ten times greater. The neutron intensity for each scan was measured by the scintillator-based Timepix3  detector \cite{Losko_2021,FumiTimepix2024}, as shown in Fig. \ref{fig:cg4b witness supp}.

The witness value calculated with these settings is found to be $S = 2.32 \pm 0.02$. The contrast decreases slightly as the spin phase is scanned, suggesting some inhomogeneity in the spin-phase coil at these currents or possibly because the total phase was too far from the MIEZE condition. This exercise indicates the known challenges of keeping the contrast high enough to obtain a witness value above the classical limit as the MIEZE frequency is increased. Examples of limiting factors in the higher frequency regimes would be maintaining adequate field homogeneity and having a sufficiently thin active detector surface.

\subsection{Initial test at the RESEDA instrument at FRM-II}

The initial measurement of this experiment was performed on the RESEDA beamline at FRM-II \cite{Franz2019}.
In that case, a velocity selector prepares the neutron beam which produces a triangular-shaped distribution of wavelengths with a bandwidth $\Delta \lambda/\lambda~=~11.6\%$ centered around 0.6~nm, in contrast to the relatively narrow 0.2\% bandwidth used in the HFIR experiment. The wider neutron wavelength bandwidth results in a smaller MIEZE envelope, which means that care must be taken to ensure that a full 2$\pi$ phase shift scan does not suppress the contrast below the minimal acceptable contrast of $1 /\sqrt{2}~\approx~71\%$.
The RESEDA field subtraction coil \cite{Jochum2020b} was used to manipulate the spin phase. In this case, the spin-phase coil was between the rf flippers, which produces the same witness as having the spin-phase coil before the rf flippers as was done in the HFIR experiments.
Furthermore, instead of the detector position, the frequency of the second rf flipper was used to manipulate the relative energy phase.
At RESEDA, neutrons are measured in a 2D CASCADE detector \cite{2010Schmidt,Kohli_2016} with an area of $200 \times 200 \text{mm}^2$; the beam used for the experiment was collimated using a series of pinholes.
RESEDA has recently been upgraded with detector translation functionality \cite{leiner2022miasans}, so
this witness test may be further measured and developed there. 

\onecolumngrid
\subsection{Picture of experimental setup on CG-4B at HFIR}
\begin{figure}[ht]
\includegraphics[width=0.99\textwidth]{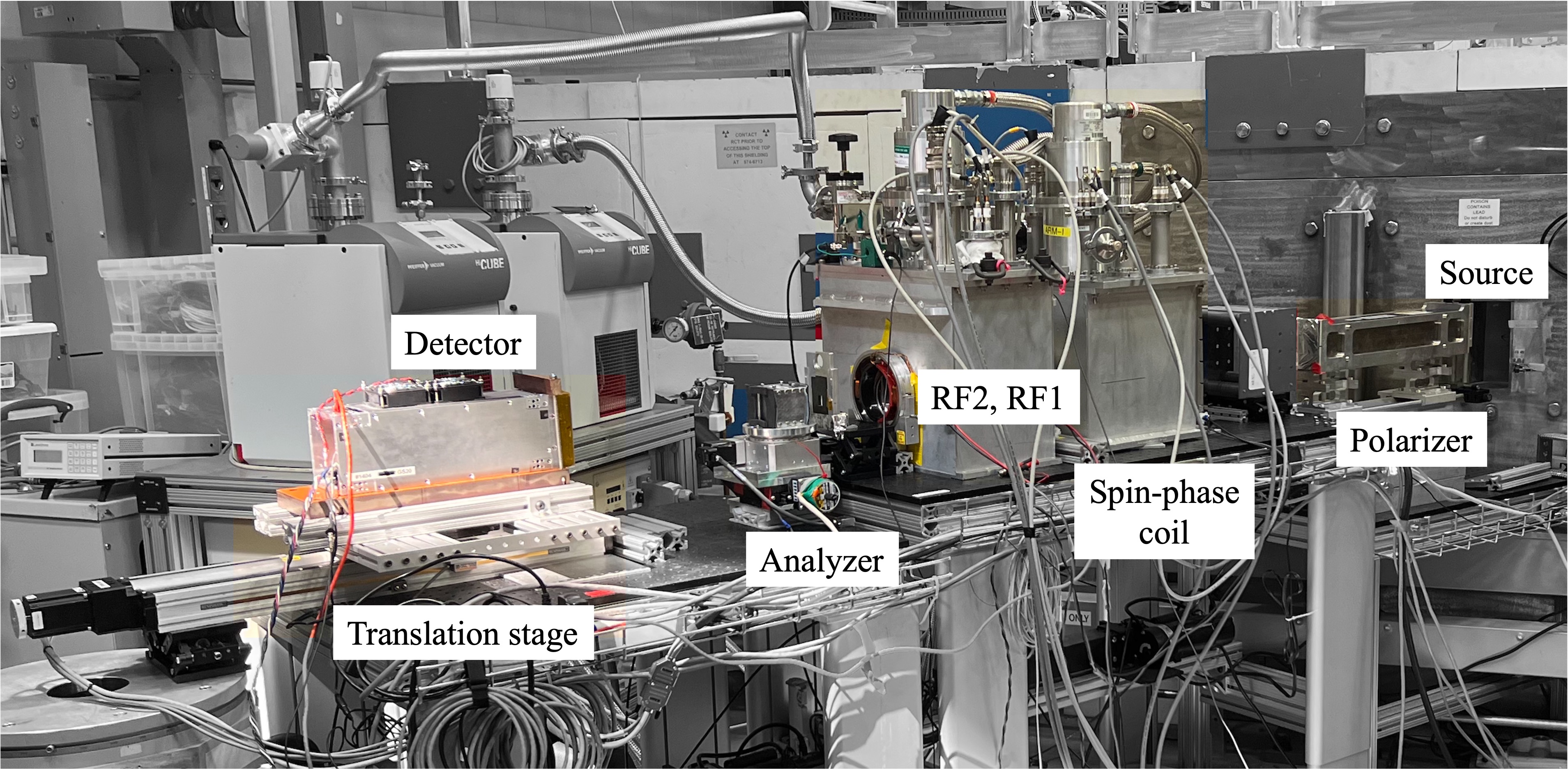}
\caption{\label{exp_photo} Photo of the components on the CG-4B beamline at HFIR diagrammed in Fig. \ref{fig:2D experiment setup} in the main text as they were when the data in Fig. \ref{fig:2dscan} was collected.}
\end{figure}

\end{document}